\documentclass[sn-mathphys-num]{sn-jnl}

\usepackage{graphicx}%
\usepackage{multirow}%
\usepackage{amsmath,amssymb,amsfonts}%
\usepackage{amsthm}%
\usepackage{mathrsfs}%
\usepackage[title]{appendix}%
\usepackage{xcolor}%
\usepackage{textcomp}%
\usepackage{manyfoot}%
\usepackage{booktabs}%
\usepackage{algorithm}%
\usepackage{algorithmicx}%
\usepackage{algpseudocode}%
\usepackage{listings}%
\usepackage{orcidlink}%

\theoremstyle{thmstyleone}%
\theoremstyle{thmstyletwo}%

\theoremstyle{thmstylethree}%

\begin{document}

\title[Shear Viscosity of an $N$-Component Gas Mixture using the Chapman--Enskog Method under Anisotropic Scatterings]{Shear Viscosity of an $N$-Component Gas Mixture using the Chapman--Enskog Method under Anisotropic Scatterings}


\author[1]{\fnm{Noah M.} \sur{MacKay}\,\orcidlink{0000-0001-6625-2321}}\email{noah.mackay@uni-potsdam.de}

\affil[1]{\orgdiv{Institute of Physics and Astronomy}, \orgname{Universit\"at Potsdam}, \orgaddress{\street{Karl-Liebknecht-Stra\ss e 24/25}, \postcode{14476} \city{Potsdam}, \country{Germany}}}


\abstract{The analytical Chapman--Enskog formula for calculating the shear viscosity $\eta$ of a relativistic ideal gas, such as a massless quark--gluon plasma, has consistently demonstrated good agreement with the numerical results obtained using the Green--Kubo relation under both isotropic and anisotropic two-body scatterings. However, past analyses of massless, multicomponent quark-gluon plasma have focused on an effective single-component ``gluon gas." The Chapman--Enskog formula for multicomponent mixtures with nonzero yet adjustable masses was previously developed for simpler cases of isotropic scatterings. This study aims to obtain the Chapman--Enskog shear viscosity formula for a massless, multicomponent mixture under general anisotropic scatterings. Since the shear viscosity depends on a linearized collision kernel, an approximation formula for the linearized collision kernel is derived under elastic and anisotropic $l+k\rightarrow l+k$ scatterings. This derived approximation agrees very well with the isotropic two-body kernels provided in previous works for both like and different species. Furthermore, for multicomponent mixtures beyond two species types, an alternative expansion method of the $N$-component Chapman--Enskog viscosity is presented. This is applied to a two-component ``binary" mixture and compared with the conventional formula for binary viscosity. The agreement between the two, for interacting and noninteracting binary mixtures, varies from moderate to well. }




\maketitle

\section{Introduction}\label{sec1}

An example of an ultrarelativistic plasma that can be directly measured and analyzed is a quark--gluon plasma (QGP) created by relativistic heavy ion collisions, such as those produced at the Relativistic Heavy Ion Collider (RHIC) and at the Large Hadron Collider (LHC) \cite{J.Adams:STAR, PHENIX:2004vcz}. In thermodynamic and statistical mechanical procedures, both analytical and numerical, the QGP is a very hot and very dense multicomponent system, i.e., multispecies system containing (anti-)quark and gluon degrees of freedom. The partons in the plasma are quasifree in an ideal thermal gas at temperatures much greater than the Hagedorn temperature: $T_H\simeq150$ MeV \cite{hage}\footnote{In this report, natural units are used, i.e., $\hbar=k_B=c=1$.}.

It is suggested by past comparisons \cite{Romatschke:2007mq, Song:2008si, ALICE:2010suc} between experimental anisotropic flow measurements and theoretical models, such as hydrodynamics, that the QGP behaves like a near-perfect fluid with a very small shear-viscosity to entropy-density ratio $\eta/s$ near the conformal string-theoretical lower bound of $1/(4\pi)$ \cite{Kovtun:2004de}. In comparison to hydrodynamic models, transport models can also describe the large amounts of observed elliptic flow in high energy heavy ion collisions once parton interactions are included \cite{Lin:2001zk, Xu:2007jv, Xu:2008av, Ferini:2008he}. In these transport models, the interactions among partons are typically represented by their interaction cross section(s), including the magnitude and angular distribution, which then determine plasma properties such as the shear viscosity $\eta$. Unlike hydrodynamic models, where the $\eta/s$ ratio (including its temperature dependence) is an input parameter, transport model calculations can only be related to the QGP shear viscosity or $\eta/s$ after applying the relation between the parton cross section(s) and $\eta$ \cite{Xu:2007jv, Ferini:2008he, Xu:2007ns, Xu:2011fi}.

The analytical analogue to numerical transport models is kinetic theory. For an elastic two-body collision $l+k\rightarrow l+k$, the collision kernel $\mathcal{C}[f]$ from the integro-differential Boltzmann equation, $p^\mu\partial_\mu f(p)=\mathcal{C}[f]$, depends on the the specific distribution function for each of the two bodies before the collision $f_{1,2}$ and after the collision $f_{3,4}$, where $f_i=f(\vec{p}_i)$; the factor $(1+af_i)$ for each distribution function, where $a=1$ for Bose--Einstein (BE), $-1$ for Fermi--Dirac (FD), and $0$ for Maxwell--Boltzmann (MB); the relative motions between the two bodies, and the differential cross section of their shared interaction. Distinctive methods for calculating the collision kernel, either simply or explicitly, bring rise to various analytical expressions for the same plasma property, such as the shear viscosity. To determine which method for expressing $\eta$ (and therewith the $\eta/s$ ratio) is more applicable for transport-modeled QGP analysis, previous studies were conducted to compare temperature-dependent analytical expressions with the numerical Green--Kubo calculations in thermal equilibrium \cite{Plumari:2012ep, MacKay:2022uxo}. In such studies, the partons in the QGP were treated as massless MB distributed particles in the numerical calculations. Therefore, the analytical methods for calculating $\mathcal{C}[f]$, which were compared with the numerical results, were deeply rooted in MB statistics.

The Chapman--Enskog (CE) method \cite{ce, degroot} for expressing the shear viscosity (and $\eta/s$) agrees well with the numerical calculations under both isotropic and anisotropic scattering cases. This enabled a secondary study in Ref. \cite{MacKay:2022uxo}, where a QGP produced by central and midcentral collisions at the LHC and at RHIC was analyzed over time. Here, the CE expressions for anisotropic $\eta$ and $\eta/s$ were semi-analytically calculated under full and partial (on-thermal but off-chemical) equilibrium cases. These calculations were plotted within the respective time frame where the plasma is in a deconfined parton state, i.e., until the system temperature reaches 150 MeV: approximately 10 fm after collision \cite{Lin:2014tya}.

In these previous studies, the QGP was simply considered to be an effective one-component system, specifically a gluon gas with added $N_f$-flavor quark degrees of freedom \cite{Plumari:2012ep, MacKay:2022uxo, Lin:2014tya, Zhang:1997ej, Wang:2021owa}. This is due to an oversaturated gluon density inside relativistic heavy ions before collision \cite{Lin:2014tya, Zhang:1997ej, Wang:2021owa}, forming a so-called ``color glass condensate.'' Thus, previous utilization of the CE shear viscosity only extends to a forward-peaking differential cross section based on the perturbative QCD (pQCD) scattering amplitude for gluon--gluon interactions (to be seen in Ref. \cite{Arnold:2003zc}, Appendix A). For a multicomponent, i.e., $N$-component mixture with nonzero particle mass, the shear viscosity has been studied for polyatomic mixtures using the Sutherland formula \cite{nasa, suth}, while for nonpolar mixtures, the CE method was used \cite{degroot, Moroz:2014}. However, these $N$-component CE viscosity studies considered the much simpler case of isotropic scatterings.

A recent study, comparing a multicomponent hydrodynamic expression with single-component kinetic transport calculations for $\eta$ -- all under isotropic scatterings --, demonstrated that a standard one-component description in general cannot be applied to a multicomponent system \cite{El:2011cp}. In light of this demonstration, a proper analysis of a multicomponent QGP would need to expand beyond the gluon gas simplification, with forward-angle scattering also taken into account. This is the motivation of this study, whereby I investigate the $N$-component CE viscosity for MB distributed ultrarelativistic particles under elastic but anisotropic scatterings. In the QGP, ultrarelativistic mechanics occur when the system temperature vastly exceeds the parton's, i.e., any flavor of (anti-)quark's rest mass. This is essential in pQCD, where the gauge coupling is small due to asymptotic freedom \cite{Gross:1973id}. This enables the scattering amplitude $|\mathcal{M}|^2$ of parton interactions to be inserted into the differential cross section, where $d\sigma/d\Omega\propto|\mathcal{M}|^2$.

Because the masses of any (anti-)quarks are neglected in an ultrarelativistic QGP, due to $T\gg m$, the different flavors of (anti-)quarks are approximated to be two generic (anti-)quarks: $q_1,~q_2,~\bar{q}_1,~\bar{q}_2$. In the kinematic sense, the interactions between different- and same-flavor (anti-)quarks are defined by their respective scattering amplitudes \cite{Arnold:2003zc}. Therefore, the measure of the relevant flavors only takes presence in the fermion degrees of freedom. To that extent, imposing a proper treatment on a QGP shear viscosity would suggest that the plasma is an effective $N=5$ mixture, composing of the partons $g,~q_1,~q_2,~\bar{q}_1,~\bar{q}_2$. Further application to this work can extend to other multicomponent, ultrarelativistic plasmas beyond collider-produced QGP. Select examples are neutron star cores, core-collapse supernovae \cite{Janka:2006fh}, and further analysis of the primordial plasma.

\section{Methods}

\subsection{The Chapman--Enskog Method, $N=1$ Anisotropic}

The 1-component Chapman--Enskog expression for $\eta$ is generally defined using nonzero parton mass and anisotropic differential cross sections \cite{Plumari:2012ep, MacKay:2022uxo, Wiranata:2012br}:
\begin{equation}\label{CE1}
\begin{split}
&\eta=\frac{T}{10}\frac{\gamma_0^2}{c_{00}},~~~\mathrm{where}~~~\gamma_0=-10\frac{K_3(z)}{K_2(z)}~~~\mathrm{and}\\
&c_{00}=\frac{16z^3}{K_2^2(z)}\int_1^\infty dy (y^2-1)^3\left[\left(y^2+\frac{1}{3z^2}\right)K_3(2zy)-\frac{y}{z}K_2(2zy) \right]\\
&~~~~~~~~~~~~~~~~~\times\int d\sigma(1-\cos^2\theta).
\end{split}
\end{equation}
In Eq. (\ref{CE1}), $K_n$ is the modified Bessel function of the second kind, and the fraction $\gamma_0$ is related to the enthalpy. Also, $z=m/T$ is a mass parameter, and $y=\sqrt{\hat{s}}/(2m)$ is the integration variable ($\hat{s}$ being the standard Mandelstam variable for the square of the center-of-momentum, or CM, energy); they couple into a mass independent variable: $2zy=\sqrt{\hat{s}}/T$. Additionally, 
\begin{equation}
\sigma_{\mathrm{tr}}=\int d\sigma(1-\cos^2\theta)
\end{equation}
 is the transport cross section \cite{Molnar:2001ux}. It is important to note that $c_{00}$ was previously referred to as the ``relativistic omega integrals'' \cite{Plumari:2012ep, Wiranata:2012br}; I will also use this name.

While the total cross section may be energy independent under a specific differential cross section $d\sigma$, the transport cross section may be energy dependent under the same $d\sigma$. This is inherently due to the $(1-\cos^2\theta)=\sin^2\theta$ angular weighting in the $\sigma_{\mathrm{tr}}$ integral; forward angle scatterings because of minimal momentum transfer (large CM energy) produces few to no measurements of the interaction. To take CM energy-driven forward-peaking angular projections into account, the transport cross section must be evaluated under a thermal average over all values of $\sqrt{\hat{s}}$ -- or alternatively under an effective thermal average with a specific integral weighting, as readily defined in Eq. (\ref{CE1}). For contextual completeness, the thermal average (using a probability density function derived under MB statistics for two massless particles \cite{Kolb:1983sj}) of the transport cross section is provided as
\begin{equation} \label{thermavg}
\langle\sigma_{\mathrm{tr}}\rangle=\frac{1}{32}\int_0^\infty du \left[u^4K_1(u)+2u^3K_2(u) \right] \int d\sigma\,\sin^2\theta,
\end{equation}
where $u=\sqrt{\hat{s}}/T$ is the integration variable, and the differential cross section is independent of the particle mass, but can have a nonzero exchange-channel screening mass.

Other analytical expressions for shear viscosity, including the hydrodynamic Israel--Stewart \cite{Israel:1963, Stewart:1971, Huovinen:2008te} and Navier--Stokes \cite{degroot} methods, and a modified version of the relaxation time approximation \cite{Plumari:2012ep}, often contain the (effectively) averaged transport cross section. This is based on physical grounds that $\eta$ depends on the momentum transfer that parton collisions generally produce. Integrating out the energy dependence via Eq. (\ref{thermavg}) or an effective thermal average of $\sigma_{\mathrm{tr}}$ allows the transport cross section to act as an anisotropic function of temperature $T$, in terms of the constant total cross section (i.e., constant coupling strength and screening mass). It is these weightings in the effective thermal averages that set various methods apart from each other, which predetermines the agreement between theoretical expectation and numerical results \cite{MacKay:2022uxo}.
 
 In the ultrarelativistic, i.e., massless case where $z\rightarrow0$ due to $T\gg m$, the ratio $c_{00}/\gamma_0^2$ in the CE expression is revised as follows \cite{MacKay:2022uxo, Plumari:2014fda}:
 \begin{equation}
\begin{split}
\lim_{z\rightarrow0}\frac{c_{00}}{\gamma_0^2}=\frac{1}{51200}&\int_0^\infty du~u^6 \left[\left(\frac{u^2}{4}+\frac{1}{3}\right)K_3(u)-\frac{u}{2}K_2(u) \right]\\
&\times\int d\sigma(1-\cos^2\theta),
\end{split}
\end{equation}
where, as in Eq. (\ref{thermavg}), the differential cross section is independent of parton mass, but there can be a nonzero screening mass.
 
 It is important to reiterate that the previous use of the CE shear viscosity was only for a 1-component system \cite{Plumari:2012ep, MacKay:2022uxo, Wiranata:2012br}. The consideration of a massive multicomponent mixture using the CE method was addressed in Refs. \cite{degroot, Moroz:2014}, while Ref. \cite{Moroz:2014} considered the specific case of isotropic scatterings. Throughout this report, I will call the methodology used in Ref. \cite{Moroz:2014} the ``Moroz framework,'' after the sole author of the original report. 

\subsection{The Moroz Framework}

The Moroz framework uses the first-order CE viscosity for $N$ species in thermal equilibrium, as defined in Refs. \cite{degroot, Moroz:2014} as
\begin{equation} \label{wk15visc}
\begin{split}
\eta=\frac{T}{10\sigma}\sum_{k=1}^{N}x_k\gamma_{0,k}C_{0,k},~~~~~\gamma_{0,k}x_k=\sum_{l=1}^N C_{0,l}C^{00}_{lk},~~~~~x_k=\frac {n_k}{\sum_{k'}^N n_{k'}}.
\end{split}
\end{equation}
In the above, $x_k$ is the so-called molar fraction, which satisfies $\sum_k^N x_k=1$; $n_k$ is the quantum number density of the $k$-th particle species, and $\sum_{k'}^N n_{k'}$ is the total number density for $N$ species. For the case of MB-distributed (anti-)quarks and gluons with spin-color degrees of freedom, their respective ``quantum'' number densities are extracted by MB statistics and factored by their respective degeneracy factors. Additionally, $C_{0,l}$ is an arbitrary factor that is dependent on the molar fraction $x_k$, the enthalpy factor $\gamma_{0,k}$, and the linearized collision kernel of a given $2\leftrightarrow2$ elastic scattering event $C^{00}_{lk}$.

For the case of $N=2$, the viscosity of a binary mixture is derived using a procedure provided in Ref. \cite{degroot}: 

1. Expand $\gamma_{0,k}x_k=\sum_{l=1}^N C_{0,l}C^{00}_{lk}$ to $N=2$: 
\begin{equation}
\begin{split}
\gamma_{0,k}x_k=C_{0,1}C^{00}_{1k}+C_{0,2}C^{00}_{2k}.
\end{split}
\end{equation}

2. Iterate for $k=1,2$, and solve for the $C_{0,k}$ terms:
\begin{equation} 
\begin{split}
&(k=1):~~~\gamma_{0,1}x_1=C_{0,1}C^{00}_{11}+C_{0,2}C^{00}_{21}\\
&~~~~~~~\rightarrow C_{0,1}=\frac{1}{C^{00}_{11}}\left(\gamma_{0,1}x_1-C_{0,2}C^{00}_{21} \right).\\[2ex]
&(k=2):~~~\gamma_{0,2}x_2=C_{0,1}C^{00}_{21}+C_{0,2}C^{00}_{22}\\
&~~~~~~~~~~~~~~~~~~~~~~~=\frac{C^{00}_{21}}{C^{00}_{11}}\left(\gamma_{0,1}x_1-C_{0,2}C^{00}_{21} \right)+C_{0,2}C^{00}_{22}\\
&~~~~~~~\rightarrow C_{0,2}=\frac{1}{(C^{00}_{11}C^{00}_{22}-(C^{00}_{21})^2)}\left(x_2\gamma_{0,2}C^{00}_{11}-x_1\gamma_{0,1}C^{00}_{21}\right).\\[2ex]
&\dot{.\hspace{.05in}.}\hspace{.4in} C_{0,1}=\frac{1}{(C^{00}_{11}C^{00}_{22}-(C^{00}_{21})^2)}\left(x_1\gamma_{0,1}C^{00}_{22}-x_2\gamma_{0,2}C^{00}_{21}\right).
\end{split}
\end{equation}

3. Expand the summation for $\eta$ to $N=2$ and utilize the $C_{0,k}$ factors obtained in Step 2 (using $\gamma_{0,1}=\gamma_{0,2}=\gamma_0$):
\begin{equation} \label{binary}
\begin{split}
\eta&=\frac{T\gamma_0}{10\sigma}\left(x_1C_{0,1}+x_2C_{0,2} \right)\\[2ex]
&=\frac{T\gamma_0^2}{10\sigma}\left(\frac{ x_1^2C^{00}_{22}-2x_1x_2C^{00}_{21}+x_2^2C^{00}_{11}}{C^{00}_{11}C^{00}_{22}-(C^{00}_{21})^2}\right).
\end{split}
\end{equation}
This $N=2$ formula is also provided in Ref. \cite{degroot, Moroz:2014}. Both references refer to the denominator as a ``determinant factor'' $\Delta_C$, as it resembles a determinant of a $2\times2$ collision kernel matrix. 

The Moroz framework is based on the Ritz variational method for linearizing the collision kernels from the Boltzmann equation \cite{degroot}. However, the Ritz method only applies to equations containing a symmetric (self-adjoint) operator, i.e., for scalar particles. As the collision operator of particles with nonzero spin is certainly nonsymmetric (see Ref. \cite{degroot}, Chap. IV, Sect 3), the Ritz method has to be put aside in favor of the more general Galerkin method \cite{galent}. As the collision kernels are symmertic operators, i.e., $C^{rs}_{lk}=C^{sr}_{kl}$ ($r$ is the spin number of species $l$ and $s$ is the spin number of species $k$), both methods yield identical results. One can expect an elastic $l_r+k_s\rightarrow l_r+k_s$ interaction to be symmetric to an elastic $k_s+l_r\rightarrow k_s+l_r$ interaction. In summary, the Ritz method can be used to linearize the kernels describing parton collisions as though the partons have no spin, i.e., $r=s=0$, as readily presented in Eqs. (\ref{wk15visc}-\ref{binary}).

Instead of keeping the scatterings generally anisotropic, isotropic scatterings were considered to provide a moderately straightforward analytical expression (using $z_l=z_k=z$):
\begin{equation} \label{wk15defs}
\begin{split}
&C_{lk}^{00}=\frac{8\sigma_{lk}}{3\sigma z^3K^2_2(z)}\Big[K_3(2z)\left(-x_lx_k(4z^2+72)+4\delta_{lk}x_l\sum_{m=1}^N x_m (4z^2+67)\right)\\
&~~~~~+\frac{1}{z}K_2(2z)\left(-x_lx_k(20z^2+16)+4\delta_{lk}x_l\sum_{m=1}^N x_m(20z^2+6)\right) \Big],\\
\end{split}
\end{equation}
where $\sigma_{lk}$ is the energy-independent $l+k\rightarrow l+k$ total cross section, and $\sigma=\sum_{l,k}\sigma_{lk}$.

\subsubsection{Pure system: $N=1$}

For a pure, 1-component gas, the CE viscosity under the Moroz framework matches the massive isotropic case found in Refs. \cite{Wiranata:2012br, Plumari:2012ep}, which used the relativistic omega integrals with $\sigma_{\mathrm{tr}}=2\sigma/3$: 
\begin{equation} \label{puregasvisc}
\begin{split}
\eta&=\frac{T}{10\sigma}\frac{\gamma_0^2}{C^{00}_{11}}~~~~~(x=1),\\
&=\frac{T}{\sigma}\left(\frac{15}{16}\frac{z^4K_3^2(z)}{(15z^2+2)K_2(2z)+(3z^3+49z)K_3(2z)} \right),
\end{split}
\end{equation}
where
\begin{equation} \label{1spec_res}
\begin{split}
&\left(C_{11}^{00}\right)_{N=1}=\frac{32}{3z^4K_2^2(z)}\left[(3z^3+49z)K_3(2z)+(15z^2+2)K_2(2z) \right],\\
&\sigma_{11}=\sigma.
\end{split}
\end{equation}

\begin{figure} [h!]
\centering
\includegraphics[width=115mm]{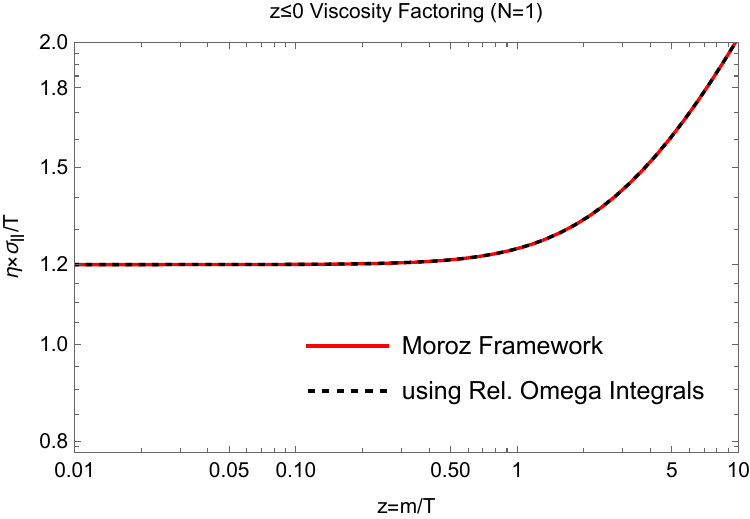}
\caption{\label{viscn1} Results from Moroz's framework (red solid) and the relativistic omega integrals (black dashed) for a pure gas $\eta\times\sigma/T$.}
\end{figure}

Figure \ref{viscn1} displays the comparison of $\eta\times\sigma/T$ between the results from the Moroz framework and those produced from the relativistic omega integrals. The two curves overlap each other, showing a perfect agreement between the two results -- more importantly showing $\eta\times\sigma/T=1.2$ for ultrarelativistic particles.

\subsubsection{Binary system: $N=2$}

For a binary mixture with $m_1=m_2=m$, we use Eq. (\ref{binary}) and the framework definitions for the like- and mixed-species kernels:
\begin{equation} \label{kernels}
\begin{split}
&C_{11}^{00}=\frac{8\sigma_{11}}{3\sigma z^3K^2_2(z)}\Big[K_3(2z)\left(-x_1^2(4z^2+72)+4(x_1^2+x_1x_2)(4z^2+67)\right)\\
&~~~~~~~~~~+\frac{1}{z}K_2(2z)\left(-x_1^2(20z^2+16)+4(x_1^2+x_1x_2)(20z^2+6)\right) \Big],\\\\
&C_{22}^{00}=\frac{8\sigma_{22}}{3\sigma z^3K^2_2(z)}\Big[K_3(2z)\left(-x_2^2(4z^2+72)+4(x_2^2+x_1x_2)(4z^2+67)\right)\\
&~~~~~~~~~~+\frac{1}{z}K_2(2z)\left(-x_2^2(20z^2+16)+4(x_2^2+x_1x_2)(20z^2+6)\right) \Big],\\\\
&C_{12}^{00}=\frac{-8x_1x_2}{3 z^3K^2_2(z)}\frac{\sigma_{12}}{\sigma}\left[K_3(2z)\left(4z^2+72\right)+\frac{1}{z}K_2(2z)\left(20z^2+16\right) \right].\\
\end{split}
\end{equation}

The cross section $\sigma=\sum_{l,k}\sigma_{lk}$ is canceled out from Eq. (\ref{binary}), leaving us with the specific interaction cross sections $\sigma_{11},~\sigma_{12},~\sigma_{22}$. Additionally, if $x_2=0$ ($n_2=0$), then $\eta$ would be for a pure gas of species 1. Likewise, if $x_1=0$, $\eta$ is for a pure gas of species 2.

The Moroz framework even demonstrates that mixed-species collisions within a mixture enhance the binary viscosity, as long as the mixture is an interacting system. Otherwise, for a noninteracting system (assuming $C^{00}_{lk}=0$ for $l\neq k$), the binary (let alone a generic $N$-component mixture) is only the summation of the pure gas contributions in the CE form, similar to the Sutherland formula provided in Refs. \cite{nasa, suth}.
 
 \section{Ritz Method under Anisotropic Scatterings} \label{ritzy}
  
Eq. (100) of Ref. \cite{degroot}, Chap. VI Sect 3c, displays the expression for the linearized collision kernel $C^{rs}_{kl}$ under MB statistics with particle mass. For $r=s=0$, letting $z_k=z_l=z$, and given the elastic collision $i+j\rightarrow k+l$, we have
\begin{equation} \label{dg100}
\begin{split}
C^{00}_{kl}&=\frac{1}{4\pi z^4K_2(2z)}\Big[x_kx_l \sum_{i,j}\left[\mathring{\overline{\pi_k^\mu \pi_k^\nu}}, \mathring{\overline{\pi_{\mu,l} \pi_{\nu,l}}}\right]_{ij|kl}\\
&~~~~~~~~+\delta_{kl}x_k\sum_{i,j,m}x_m\left[\mathring{\overline{\pi_k^\mu \pi_k^\nu}}, \mathring{\overline{\pi_{\mu,k} \pi_{\nu,k}}}\right]_{ij|km}\\
&~~~~~~~~~~~~~~~~ -2x_k\sum_{j,m} x_m \left[\mathring{\overline{\pi_k^\mu \pi_k^\nu}}, \mathring{\overline{\pi_{\mu,l} \pi_{\nu,l}}}\right]_{kj|lm}\Big],
\end{split}
\end{equation}
where, using the $\hat{s}$-Mandelstam variable,
\begin{equation} \label{bef}
\begin{split}
[F_k,G_l]_{ij|kl}&:=\frac{1}{2\sigma(4\pi T^3)^2z^4K_2^2(z)}\\
&~~~~~~~~~~\times\int \frac{d^3\vec{p}_i}{p^0_i}\frac{d^3\vec{p}_j}{p^0_j}\frac{d^3\vec{p}_k}{p^0_k}\frac{d^3\vec{p}_l}{p^0_l} \exp(-\tau_k-\tau_l)F_kG_l\\
&~~~~~~~~~~~~~~~~\times \hat{s}\sigma_{ij|kl}(\hat{s},\theta)\delta^{(4)}(P_i+P_j-P_k-P_l)
\end{split}
\end{equation}
\begin{equation} \label{aft}
\begin{split}
&\Rightarrow [F_k,G_l]_{kl|kl}:=\frac{\pi}{\sigma(4\pi T^3)^2z^4K_2^2(z)}\\
&~~~~~~~~~~~~~~~~~~~~~~~~~~\times\int \frac{d^3\vec{p}_k}{p^0_k}\frac{d^3\vec{p}_l}{p^0_l} \exp(-\tau_k-\tau_l)F_kG_l\\
&~~~~~~~~~~~~~~~~~~~~~~~~~~\times\int_{4m^2}^\infty d\hat{s}\,\delta(P^2-\hat{s}) \sigma_{kl|kl}(\hat{s},\theta)\sqrt{\hat{s}(\hat{s}-4m^2)},
\end{split}
\end{equation}
$F_k$ and $G_l$ are arbitrary functions of the 3-momenta $\vec{p}_{k}$ and $\vec{p}_{l}$ respectively, $\tau_n=|\vec{p}_n|/T$, and $P_n=(p^0_n,\vec{p}_n)$ is the 4-momentum. The reduction from Eq. (\ref{bef}) to Eq. (\ref{aft}) is based on the conservation of energy and 3-momentum, which is illustrated by the total 4-momenta under elastic scatterings: $P_i+P_j=P_k+P_l\equiv P$. This effect considers a purely elastic $k+l\rightarrow k+l$ interaction, which Moroz previously considered \cite{Moroz:2014}.

The tensoral collision brackets in Eq. (\ref{dg100}) are reduced down to collision brackets containing powers of $\tau_n$ and a linear combination of integral functions (c.f. Ref. \cite{degroot}, pg. 389, Eq. (52)):
\begin{equation}
\begin{split}
&\left[\mathring{\overline{\pi_k^\mu \pi_k^\nu}}, \mathring{\overline{\pi_{\mu,l} \pi_{\nu,l}}}\right]_{kl|kl}=-\frac{4}{3}\left[\tau_k^2, \tau_l^2\right]_{kl|kl}+\frac{z^2}{3}\left[1, \tau_l^2\right]_{kl|kl}\\
&~~~~~~~~~~+\frac{z^2}{3}\left[\tau_k^2,1\right]_{kl|kl}-\frac{z^4}{3}\left[1,1\right]_{kl|kl}-2\left[\tau_k\,\overline{\pi_k^\mu}, \tau_l\,\overline{\pi_{\mu,l}}\right]_{kl|kl}\\
&~~~~~~~~~~+\frac{1}{16}\left(J_{kl|kl}^{(2,0,0,0,0|4,0)}-2J_{kl|kl}^{(1,0,0,0,1|2,0)}+J_{kl|kl}^{(0,0,0,0,2|0,0)} \right).
\end{split}
\end{equation}
Here (c.f. Ref. \cite{degroot}, pgs. 386-387, Eqs. (41) and (43)),
\begin{equation} \label{jint}
\begin{split}
&J_{kl|kl}^{(a,0,0,0,f|q,0)}=\frac{(2-\delta_{kl})\pi}{\sigma z^2K_2^2(z)}~(2z)^{2(a+f)+3}\\
&~~~~~\times\int_1^\infty dy~y^{2(a+1)}(y^2-1)^{f+1} K_1(2zy)\int_{-1}^1 d\chi\left[1-\chi^f\right]\sigma_{kl|kl}(2zy,\chi),
\end{split}
\end{equation}
where $y=\sqrt{\hat{s}}/(2m)$ and $\chi=\cos\theta$ are the integration variables, and  $\sigma_{kl|kl}(2zy,\chi)=\sigma_{kl|kl}(\hat{s},\theta)$. The superscript factor $q$ plays no role in Eq. (\ref{jint}), yet $f=0$ leads to $J_{ij|kl}^{(a,0,0,0,f|q,0)}=0$. 
 
To limit the redundancy of the $kl|kl$ subscripts to denote the $k+l\rightarrow k+l$ interaction, I will only write $kl$ with the differential cross section.
 
 \subsection{Massless Partons, $z\rightarrow0$}

 For $z\rightarrow0$, the tensoral collision brackets can be written as a simpler expression of integral functions:
\begin{equation}
\begin{split}
&\left[\mathring{\overline{\pi_k^\mu \pi_k^\nu}}, \mathring{\overline{\pi_{\mu,l} \pi_{\nu,l}}}\right]=-\frac{4}{3}\left[\tau_k^2, \tau_l^2\right]-2\left[\tau_k\,\overline{\pi_k^\mu}, \tau_l\,\overline{\pi_{\mu,l}}\right]\\
&~~~~~~~~~~~~~~~~~~~~~~~~~~~~~~+\frac{1}{16}\left(-2J^{(1,0,0,0,1|2,0)}+J^{(0,0,0,0,2|0,0)} \right)\\
&~~~~~~~~~~~~~~~~~~~~~\equiv T_{kl}+J_{kl}.
\end{split}
\end{equation}
Here, I call
\begin{equation} \label{tint}
\begin{split}
T_{kl}&\equiv-\frac{4}{3}\left[\tau_k^2, \tau_l^2\right]-2\left[\tau_k\,\overline{\pi_k^\mu}, \tau_l\,\overline{\pi_{\mu,l}}\right]\\
&=\frac{-2\pi}{\sigma(4\pi T^3)^2}\int \frac{d^3p_k}{p^0_k}\frac{d^3p_l}{p^0_l} \exp(-\tau_k-\tau_l)\tau_k\tau_l\left(\frac{2}{3}\tau_k\tau_l+\overline{\pi_k^\mu}\overline{\pi_{l,\mu}} \right)\\
&~~~~~~~~~~~~~~~~~~~~~~\times\int_0^\infty d\hat{s}\,\delta(P^2-\hat{s}) \hat{s}\sigma_{kl}(\hat{s},\theta),
\end{split}
\end{equation}
and
\begin{equation} \label{jint}
\begin{split}
J_{kl}&\equiv\frac{1}{16}\left(-2J^{(1,0,0,0,1|2,0)}+J^{(0,0,0,0,2|0,0)} \right)\\
&=\frac{(\delta_{kl}-2)\pi}{64\sigma}\int_0^\infty d\hat{s}~\frac{\hat{s}^4}{2T^{10}}K_1(\sqrt{\hat{s}}/T)\int_{-1}^1 d\chi~\frac{1}{4}(1-2\chi+\chi^2)~\sigma_{kl}(\hat{s},\chi).
\end{split}
\end{equation}

 In the first integral function $T_{kl}$, defined as Eq. (\ref{tint}), one can use Eq. (46) of Ref. \cite{degroot}, pg. 387, to define the vector product $\overline{\pi_k^\mu}\,\overline{\pi_{\mu,l}}$ as a so-called ``contraction.'' Here, it is defined based on the Mandelstam variables $\hat{s}$ and $\hat{t}$:
\begin{equation}\label{contra}
\overline{\pi_k^\mu}\,\overline{\pi_{\mu,l}}=\frac{1}{4T^2}\left[(1+\alpha_{kl})^2\hat{s}+\hat{t}\,\right]-\tau_k\tau_l,
\end{equation}
where $\alpha_{lk}=(m_k^2-m_l^2)/\hat{s}$, which is zero for massless particles, and $\hat{t}$ is the Mandelstam variable for the square of the momentum transfer.  For $z\rightarrow0$,
\begin{equation}
\begin{split}
T_{kl}&=\frac{-2\pi}{\sigma(4\pi T^3)^2}\int \frac{d^3p_k}{p^0_k}\frac{d^3p_l}{p^0_l} \exp(-\tau_k-\tau_l)\tau_k\tau_l\left(-\frac{1}{3}\tau_k\tau_l+\frac{1}{4T^2}\left[\hat{s}+\hat{t}\,\right] \right)\\
&~~~~~~~~~~~~~~~~~~~~~~\times\int_0^\infty d\hat{s}\,\delta(P^2-\hat{s}) \hat{s}\sigma_{kl}(\hat{s},\theta)
\end{split}
\end{equation}
\begin{equation}
\begin{split}
\Rightarrow T_{kl}&=\frac{-\pi}{128\sigma T^9}\int_0^\infty {d\hat{s}}~\hat{s}^{7/2}K_1\left(\sqrt{\hat{s}}/T\right) \int_{-\hat{s}}^0 d\hat{t}\left(\frac{8}{3}+\frac{4\hat{t}}{\hat{s}}\right)\frac{d\sigma_{kl}}{d\hat{t}},
\end{split}
\end{equation}
where $d\sigma_{kl}/d\hat{t}$ is the differential cross section based on momentum transfer. The $\int d\hat{t}[\dots]$ integral does not define the transport cross section, as the integral evaluates to $8\sigma_{kl}/3$ for forward-peaking scatterings (as long as $\sigma_{kl}$ is energy- independent). However, it does evaluate to $2\sigma_{kl}/3$ under isotropic scatterings, where $d\sigma_{kl}/d\hat{t}=\sigma_{kl}/\hat{s}$.

In the second integral function $J_{kl}$, defined as Eq. (\ref{jint}), the $\int d\chi[\dots]$ integral indeed defines the transport cross section, which suggests that the relativistic omega integrals $c_{00}$ will originate from the $J_{kl}$ integrals after the summation over all particle indices $i,j,m$ (as shown in Eq. [\ref{dg100}]). Therefore,
\begin{equation}
\begin{split}
\Big[\mathring{\overline{\pi_k^\mu \pi_k^\nu}}, &\mathring{\overline{\pi_{\mu,l} \pi_{\nu,l}}}\Big]=\frac{\pi}{64\sigma} \Big\{-\frac{1}{2}\int_0^\infty d\hat{s}~\frac{\hat{s}^{7/2}}{2T^9}K_1\left(\sqrt{\hat{s}}/T\right) \int_{-s}^0 d\hat{t}\left(\frac{8}{3}+\frac{4\hat{t}}{\hat{s}}\right)\frac{d\sigma_{kl}}{d\hat{t}}\\
&+(\delta_{kl}-2)\int_0^\infty d\hat{s}~\frac{\hat{s}^{4}}{2T^{10}}K_1\left(\sqrt{\hat{s}}/T\right) \int_{-s}^0 d\hat{t}\left(\frac{-4\hat{t}^2}{\hat{s}^2}-\frac{4\hat{t}}{\hat{s}}\right)\frac{d\sigma_{kl}}{d\hat{t}}  \Big\}, 
\end{split}
\end{equation}
where the transport cross section in the $J_{kl}$ integral function is defined in terms of the $\hat{s}$ and $\hat{t}$ Mandelstam variables.

As there are three tensoral collision brackets in Eq. (\ref{dg100}) with $z\rightarrow0$, two of which depend on the index for total species $m\in[1,N]$, I propose a brute-force approximation to the collision kernel in such a way that $J_{kl}$ -- summed over $i,j,m$ -- would define the relativistic omega integrals for like-species (such that the general $N=1$ results will be preserved), and that $T_{kl}$ would only be present for mixed-species interactions, i.e., vanish under like-species interactions. This brute-force approximation for $C^{00}_{kl}$, using the mass-independent integration variable $u=\sqrt{\hat{s}}/T$, is presented as follows: 
\begin{equation} \label{anisodg}
\begin{split}
C^{00}_{kl}&\simeq\frac{5^{\delta_{N,1}}}{32z^2\sigma}\frac{\left[\pi^{(\delta_{lk}-1)}2^{(2-\delta_{lk})}\right]^{\Delta_{N,1}}}{(11-6\delta_{kl})}\left[x_kx_l-2\delta_{kl}x_k\left(\frac{7}{4}\right)^{\delta_{kl}\Delta_{N,1}}\sum_{m=1}^Nx_m \right]\\
&~~~\times\Big\{\frac{(\delta_{kl}-1)}{2}\int_0^\infty du~u^8K_1(u)\int_{-u^2T^2}^0 d\hat{t}\left(\frac{8}{3}+\frac{4\hat{t}}{u^2T^2}\right)\frac{d\sigma_{kl}}{d\hat{t}}\\
&~~~~~~~+(\delta_{kl}-2)\int_0^\infty du~u^6 \left[\left(\frac{u^2}{4}+\frac{1}{3}\right)K_3(u)-\frac{u}{2}K_2(u) \right]\\
&~~~~~~~~~~~~~~~~~~~~~~~~~~\times\int_{-u^2T^2}^0 d\hat{t}\left(\frac{-4\hat{t}^2}{u^4T^4}-\frac{4\hat{t}}{u^2T^2}\right)\frac{d\sigma_{kl}}{d\hat{t}}   \Big\} .
\end{split}
\end{equation}
Here, $\Delta_{N,1}=1-\delta_{N,1}$, where $\delta_{N,1}$ is the Kronecker delta between the total number of species and the case of $N=1$. In other words, $\Delta_{N,1}=0$ for one-component systems, and $\Delta_{N,1}=1$ for multicomponent systems. Note that for like-species, the first integral vanishes from the collision kernel.

The relativistic omega integrals are explicitly expressed in Eq. (\ref{anisodg}); the approximation of $C^{00}_{kl}$ would provide identical isotropic and anisotropic results for $N=1$, where
\begin{equation}
\begin{split}
\left(C^{00}_{11}\right)_{N=1}=\frac{1}{32z^2\sigma_{11}}&\int_0^\infty du~u^6 \left[\left(\frac{u^2}{4}+\frac{1}{3}\right)K_3(u)-\frac{u}{2}K_2(u) \right]\\
&\times\int_{-u^2T^2}^0 d\hat{t}\left(\frac{-4\hat{t}^2}{u^4T^4}-\frac{4\hat{t}}{u^2T^2}\right)\frac{d\sigma_{11}}{d\hat{t}}.
\end{split}
\end{equation}

\subsubsection{Isotropic $N=2$} \label{geniso}

For the question of consistency, we compare Eq. (\ref{anisodg}) with the $z\rightarrow0$ Moroz framework for the case of isotropic $N=2$, with the analytical expressions listed in Table \ref{tab1}.  \\

\begin{table}[h]
\caption{Analytical expressions of $C^{00}_{lk}$ for $N=2$ and $z\rightarrow0$}\label{tab1}%
\begin{tabular}{lll}
\toprule
	 & Approximation & Moroz Framework   \\
\midrule
$C^{00}_{11}\times\sigma z^2$ & ${133.333}\sigma_{11} \left[x_1^2 +1.4x_1x_2\right]$ &   ${133.333}\sigma_{11} \left[x_1^2 +1.4x_1x_2\right]$  \\\\
$ C^{00}_{22}\times\sigma z^2$ & ${133.333}\sigma_{22} \left[x_2^2 +1.4x_2x_1\right]$ & ${133.333}\sigma_{22} \left[x_2^2 +1.4x_2x_1\right]$  \\\\
 $C^{00}_{12}\times\sigma z^2$ & ${53.0902}\sigma_{12} \left[-x_1x_2 \right]$ & ${53.3333}\sigma_{12} \left[-x_1x_2 \right]$  \\
\botrule
\end{tabular}
\end{table}

The closeness between the approximated and the Moroz $C^{00}_{12}$ provides a 1:1 ratio of 1.00458. For the like-species kernels, the approximation and the Moroz framework results are identical. Therefore, Eq. (\ref{anisodg}) can be used for $N\geq2$ anisotropic scatterings. 

\section{An Alternative, Iterative $N\geq2$ CE Viscosity} \label{iterative}

In this section, I propose an alternative procedure to determine the CE viscosity of an $N$-component system. It is nevertheless based on the procedure provided by Ref. \cite{degroot}, especially for $N=2$, but it can be generalized for larger $N$. The motivation is to address the inherent complexity of solving for the determinant factor $\Delta_C$ for $N>2$, especially for $N=5$ for the gluons and (anti-)quarks in a QGP.

Providing $N=2$ as a leading example, and eventually comparing it with Eq. (\ref{binary}),

1. Iterate $\gamma_{0,k}x_k=\sum_{l=1}^N C_{0,l}C^{00}_{lk}$ for $N=1,2$: 
\begin{equation}
\begin{split}
\gamma_{0,k}x_k&=C_{0,1}C^{00}_{1k}~~~~~~~~~~~~~~~~~~~~(N=1)\\[2ex]
&=C_{0,1}C^{00}_{1k}+C_{0,2}C^{00}_{2k}~~~~~~(N=2).
\end{split}
\end{equation}

2. For $N=1$, solve for $C_{0,1}$ in terms of $\gamma_{0,1},~x_1,$ and $C^{00}_{11}$ (where $k=1$ for one species in the viscosity summation):
\begin{equation}
\begin{split}
\gamma_{0,1}x_1=C_{0,1}C^{00}_{11}~~~\rightarrow~~~ C_{0,1}=\frac{\gamma_{0,1}x_1}{C^{00}_{11}}.
\end{split}
\end{equation}

3. For $N=2$, solve for $C_{0,2}$ in terms $C^{00}_{2k},~C^{00}_{1k},~\gamma_{0,k},~x_k$, and the expression for $C_{0,1}$ previously obtained in Step 2:
\begin{equation}
\begin{split}
&\gamma_{0,k}x_k=C_{0,1}C^{00}_{1k}+C_{0,2}C^{00}_{2k}\\
&\rightarrow~~~C_{0,2}=\left(\gamma_{0,k}x_{k}-\frac{\gamma_{0,1}x_1C^{00}_{1k}}{C^{00}_{11}} \right)\frac{1}{C^{00}_{2k}}.
\end{split}
\end{equation}
Then iterate for $k=1,2$:
\begin{equation} \label{errat}
\begin{split}
&(k=1):~~~C^{(1)}_{0,2}=\left(\gamma_{0,1}x_{1}-\frac{\gamma_{0,1}x_1C^{00}_{11}}{C^{00}_{11}} \right)\frac{1}{(C^{00}_{21}+C^{00}_{12})}=0\\[2ex]
&(k=2):~~~C^{(2)}_{0,2}=\left[\gamma_{0,2}x_{2}-\frac{\gamma_{0,1}x_1}{C^{00}_{11}}\left(C^{00}_{12}+C^{00}_{21}\right) \right]\frac{1}{C^{00}_{22}}.
\end{split}
\end{equation}
Note that for $l\neq k$, the subscripts of the collision kernel were permuted, and the different permutations were added. However, as there is a symmetric relation for the collision kernels, $C^{00}_{12}=C^{00}_{21}$. Add together the iterations to have $C_{0,2}$:
\begin{equation}
\begin{split}
C_{0,2}=C^{(1)}_{0,2}+C^{(2)}_{0,2}=\left[\gamma_{0,2}x_{2}-\gamma_{0,1}x_1\frac{2C^{00}_{12}}{C^{00}_{11}} \right]\frac{1}{C^{00}_{22}}.
\end{split}
\end{equation}

4. Expand the summation for $\eta$ to $N=2$ and utilize the $C_{0,k}$ factors obtained in Steps 2-3 to obtain a binary CE viscosity (using $\gamma_{0,1}=\gamma_{0,2}=\gamma_0$):
\begin{equation} \label{newbicrypt}
\begin{split}
\eta&=\frac{T}{10\sigma}\left(x_1\gamma_{0,1}C_{0,1}+x_2\gamma_{0,2}C_{0,2} \right)\\[2ex]
&=\frac{T\gamma_0^2}{10\sigma}\left( \frac{x_1^2}{C^{00}_{11}}+\frac{x_2^2}{C^{00}_{22}}-\frac{2x_1x_2C^{00}_{12}}{C^{00}_{11}C^{00}_{22}} \right).
\end{split}
\end{equation}

The benefit of using this alternative expression is having the clear distinction between the ``pure gas'' and the mixed-species viscosities. This is reminiscent of a zero-mass redefinition of the Sutherland formula for the $N$-component shear viscosity, which can be defined as the direct sum of all individual ``pure gas'' viscosities: $\eta=\sum_i \eta_{ii}$ \cite{nasa}. However, for an interacting mixture, mixed-species interference contributes to the enhancement of a $N$-component viscosity.

\subsection{Application: An $N=2$ QGP with partons $g$ and $q$}

I will use the QGP -- interacting ($C_{gq}^{00}\neq0$) and noninteracting ($C_{gq}^{00}=0$) -- as an example to show how the quark (using $N_f\leq3$ for the up, down and strange flavors) and gluon number densities affect the binary viscosity. For an effective comparison between Eqs. (\ref{binary}) -- using the Moroz isotropic kernels provided as Eq. (\ref{wk15defs}) -- and (\ref{newbicrypt}) -- using the massless, anisotropic approximation of $C_{lk}^{00}$ offered in Eq. (\ref{anisodg}) --, we would have to consider isotropic scatterings. This would be a flat line over all $z$ for Eq. (\ref{newbicrypt}) via Eq. (\ref{anisodg}), so the true comparison between the Moroz and iterated CE methods lies within the $z\rightarrow0$ region.

 For a QGP, the molar fractions are (using a MB statistical number density)
\begin{equation} \label{wk15gqx}
\begin{split}
x_g=\frac{n_g}{n_g+n_q},~~~x_q=\frac{n_q}{n_g+n_q},~~~\mathrm{where}~~~n_j=d_j\frac{T^3}{\pi^2},
\end{split}
\end{equation}
and $d_g=16$ and $d_q=6N_f$ are the gluon and quark degeneracy factors, respectively. Therefore, $n_g+n_q\propto 2(8+3N_f)$ for the sum of the two number densities. Essentially, $x_g$ and $x_q$ would only be the ratio between the degeneracies of the specific species and of the total QGP. Note that the flavor number $N_f$ also contributes to the species enhancement of the gas, even though the flavors of quark are approximated to be one generic quark due to $T\gg m$. Nevertheless, the $N_f=1$ calculations would be most reliable for a proper binary QGP. 

It is important to recognize that quark flavor affects the number of species in the QGP, by switching on and off the quark population. If there are no quark flavors in the QGP degrees of freedom, i.e., $N_f=0$, the QGP is a gluon gas with $N=1$. However, with nonzero $N_f$, the total number of species also consists of the relevant quark types. For this case of $N=2$ with gluons and arbitrary quarks:
\begin{equation}
N_{\mathrm{QGP}}=2-\delta_{0,N_f},
\end{equation}
where $\delta_{0,N_f}$ is the Kronecker delta between no quark flavors and a specific $N_f$. This expression can be generally defined for any species number $N$:
\begin{equation}
\begin{split}
\rightarrow N_{\mathrm{QGP}}= N(1-\delta_{0,N_f})+\delta_{0,N_f}.
\end{split}
\end{equation}

 For the gluon--gluon, same flavor quark--quark, and quark--gluon interactions, the respective total cross sections, derived by ``large ${\hat{s}}$'' differential cross sections based on the pQCD $|\mathcal{M}|^2/g^4$ scattering amplitudes (c.f. Appendix A in Ref. \cite{Arnold:2003zc}), are energy independent:
\begin{equation} \label{wk15sigs}
\begin{split}
\sigma_{gg}=\frac{9g^4}{32\pi m_D^2},~~~\sigma_{qg}=\frac{g^4}{8\pi m_D^2},~~~\sigma_{qq}=\frac{g^4}{18\pi m_D^2}.
\end{split}
\end{equation}
All total cross sections are proportional to $g^4/m_D^2$, where $g$ is the gauge coupling and $m_D$ is the screening mass. Both $g$ and $m_D$ are running parameters, whose ratio $g^4/m_D^2$ reduces the total cross section with an increase in temperature via asymptotic freedom. Rather than plotting $\eta\times\sigma/T$ vs. $z$ for a noninteracting and interacting binary viscosity, it is better to plot $\eta\times g^4/(m_D^2T)$ vs. $z$ to remove any inherent temperature dependence.

\subsubsection{Noninteracting}

\begin{figure}[h!]
\centering
\includegraphics[width=120mm]{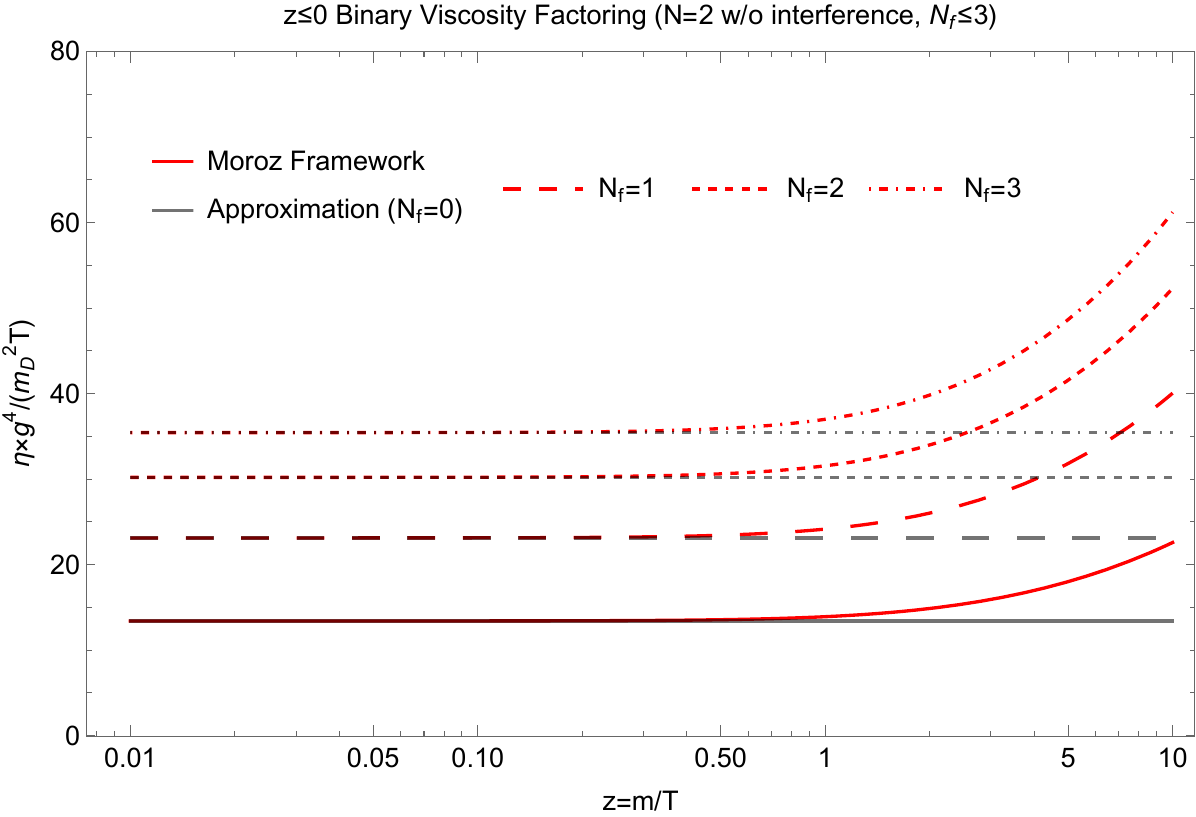}
\caption{\label{viscn2_nonint1} $\eta\times g^4/(m_D^2T)$ of a noninteracting (no cross-species interference) binary mixture. The solid curves demonstrate the gluon gas results ($N_f=0$) from the Moroz (red) and Approximation (gray) frameworks. Larger flavor numbers $N_f=1$ (long-dashed), 2 (small-dashed), and 3 (dot-dashed) are also compared; an increase in the flavor number enhances the curves.}
\end{figure}

For a noninteracting QGP ($C^{00}_{qg}=0$), both Eqs. (\ref{binary}) and (\ref{newbicrypt}) are defined as a linear combination of the two ``pure gas'' viscosities:
\begin{equation}
\eta=\frac{T\gamma_0^2}{10\sigma}\left(\frac{x_g^2}{C^{00}_{gg}}+\frac{x_q^2}{C^{00}_{qq}} \right).
\end{equation}
The subtle difference is the evaluation of $C_{ll}^{00}$ via the Moroz framework and approximation formula, which were proven to have a perfect 1:1 ratio in Section \ref{geniso}, and again demonstrated in Table \ref{tab2}. Figure \ref{viscn2_nonint1} displays the total $\eta\times g^4/(m_D^2T)$ versus $z$ between the two methods with no cross-species interference. $N_f=0$ represents the pure gluon viscosity result; with increasing $N_f$ the noninteracting viscosity is enhanced by the number of quark flavors in the QGP degrees of freedom.

\begin{table}[h]
\caption{Noninteracting $\eta\times g^4/(m_D^2T)$ for a $N=2$ QGP}\label{tab2}
\begin{tabular}{llll}
\toprule
{} & Moroz  ($z\rightarrow0$) & Approximation  & 1:1 Ratio   \\
\midrule
 $N_f=0$ & 13.4153 & 13.4068 & 1.00063 \\
 
$N_f=1$ & 23.1259 & 23.1259 & 1.00000  \\
 
$N_f=2$ & 30.2101 & 30.2101 & 1.00000 \\
 
$N_f=3$ & 35.4394 & 35.4394 & 1.00000 \\
\botrule
\end{tabular}
\end{table}

\subsubsection{Interacting}

For an interacting QGP ($C^{00}_{qg}\neq0$), we take the explicit definitions of the ``Full Moroz'' (Eq. [\ref{binary}] using Eq. [\ref{wk15defs}]), and the ``Iterative Approximation'' (Eq. [\ref{newbicrypt}] using Eq. [\ref{anisodg}]) formulas.

\begin{figure}[h!]
\centering
\includegraphics[width=120mm]{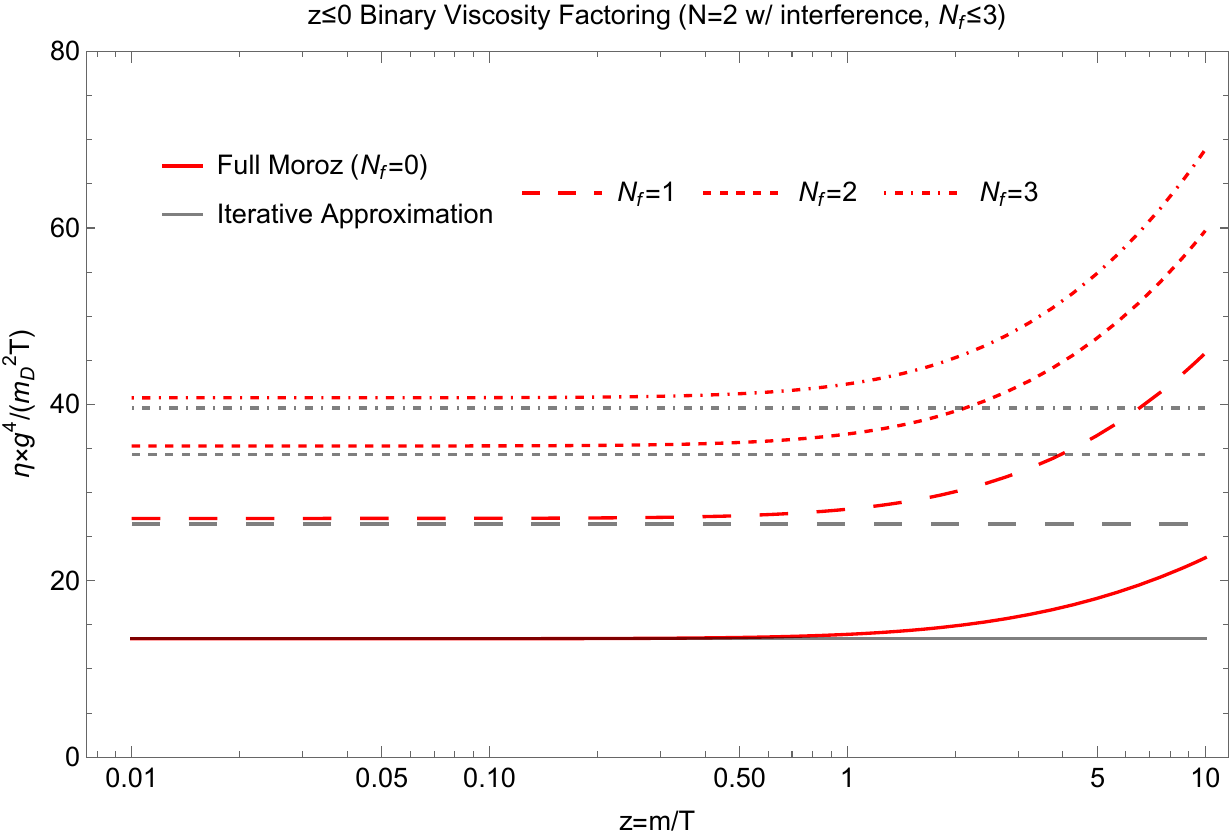}
\caption{\label{viscn2_int1} $\eta\times g^4/(m_D^2T)$ of an interacting (including cross-species interference) binary mixture. Like in Figure \ref{viscn2_nonint1}, the solid curves demonstrate the gluon gas results ($N_f=0$) from the ``Full Moroz'' (red) and iterative approximation (gray) frameworks. Larger flavor numbers $N_f=1$ (long-dashed), 2 (small-dashed), and 3 (dot-dashed) were also compared. }
\end{figure}

 Figure \ref{viscn2_int1} shows the total $\eta\times g^4/(m_D^2T)$ versus $z$ between the two methods, however with cross-species interference included. As like for the noninteracting viscosity, $N_f=0$ and $z\rightarrow0$ indicates a massless gluon gas, for which both methods agree with a 1:1 ratio of 1.00016. For larger $N_f$ under $z\rightarrow0$, the results from Full Moroz are slightly greater than those from Iterative Approximation, at average with a 1:1 ratio of $\sim0.97$; the exact ratios are listed in Table \ref{tab3}. This discrepancy is due to the inherent calculation differences between Eqs. (\ref{binary}) and (\ref{newbicrypt}). 
 
\begin{table}[h]
\caption{Interacting $\eta\times g^4/(m_D^2T)$ for a $N=2$ QGP}\label{tab3}
\begin{tabular}{llll}
\toprule
{} & Full Moroz ($z\rightarrow0$) & Iterative Approximation  & 1:1 Ratio  \\
\midrule
 $N_f=0$ & 13.4223 & 13.4245 & 1.00016 \\
 
 $N_f=1$ & 27.0684 & 26.4532 & 0.97727  \\
 
$N_f=2$ & 35.2762 & 34.2971 & 0.97224  \\
 
$N_f=3$ & 40.7420 & 39.5951 & 0.97185  \\
\botrule
\end{tabular}
\end{table}

\begin{figure}[h!]
\centering
\includegraphics[width=120mm]{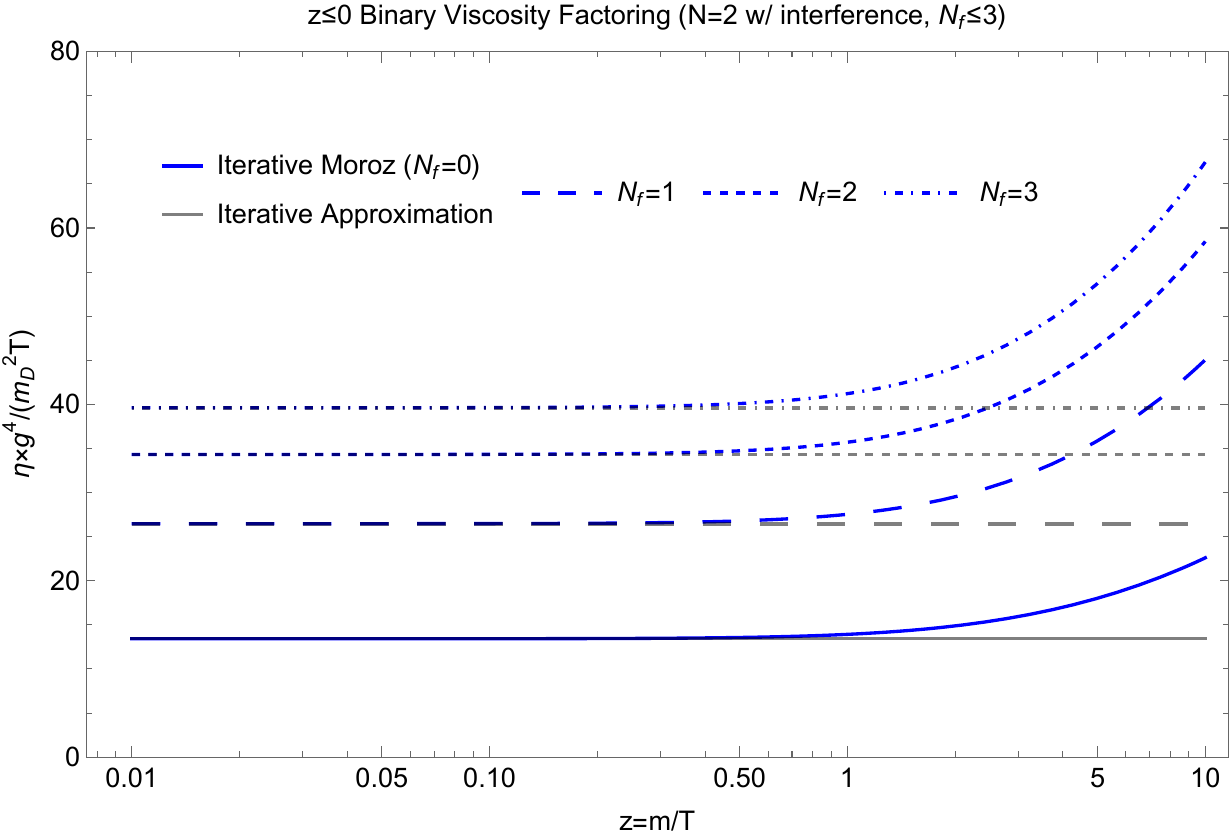}
\caption{\label{viscn2_int2} Second plot of $\eta\times g^4/(m_D^2T)$ for an interacting binary mixture. As in Figure \ref{viscn2_int1}, the curve patterns denote different numbers of relevant quark flavors, where the comparison between the iterative Moroz (blue) and iterative approximation (gray) methods is shown. }
\end{figure}

Now, the kernels from the Moroz framework are utilized in the iterative CE formula. The objective is to compute a new comparison between Eqs. (\ref{wk15defs}) and (\ref{anisodg}), only providing Eq. (\ref{newbicrypt}) as the formula for viscosity. The combination of Eqs. (\ref{wk15defs}) and (\ref{newbicrypt}) displays an ``Iterative Moroz'' method, which agrees well with the iterative approximation method with 1:1 ratios within $0.999\sim1.00$, as shown in Figure \ref{viscn2_int2} and Table \ref{tab4}. 

\begin{table}[h]
\caption{Interacting $\eta\times g^4/(m_D^2T)$ for a $N=2$ QGP, Take 2}\label{tab4}
\begin{tabular}{llll}
\toprule
{} & Iterative Moroz ($z\rightarrow0$) & Iterative Approximation  & 1:1 Ratio  \\
\midrule
 $N_f=0$ & 13.4217 & 13.4245 & 1.00020 \\
 
 $N_f=1$ & 26.4684 & 26.4532 & 0.99942  \\
 
$N_f=2$ & 34.3158 & 34.2971 & 0.99946  \\
 
$N_f=3$ & 39.6141 & 39.5951 & 0.99952 \\
\botrule
\end{tabular}
\end{table}

\section{Discussion}

The collective motivation of Sections \ref{ritzy} and \ref{iterative} is to expand beyond dependency on the Moroz framework. While the Moroz framework considers $N$-component mixtures containing particles with mass (which can be adjusted to be very small), the framework depends on isotropic scatterings. Anisotropic scatterings, using any differential cross section that computes an energy independent total cross section, is a more generalized case for a multicomponent mixture, which includes isotropic scattering as a specific and accessible scenario.

Accessibility to the isotropic scenario via the obtained approximation formula (Eq. [\ref{anisodg}]), and comparison with the massless Moroz framework for 1- and 2-component mixtures was essentially a sanity check to test for consistency and therefore form a basis of trust to expand into mixtures beyond two species types. Applying the formulas to an $N=2$ QGP containing gluons and same-flavor quarks, the 1-component comparison (using gluon degrees of freedom) and 2-component comparison (using gluon and quark degrees of freedom up to three flavors) results are in good agreement. Expansion beyond $N=2$ is intended to consider other prominent parton types in the QGP: different flavor quarks, and anti-quarks.

\subsection{CE iterative approximation: $N>2$ expansions}

Continuing the procedure provided in Section \ref{iterative}, we can define the factors $C_{0,3},~C_{0,4},$ and $C_{0,5}$, which are essential for the summation expansion of $\eta$ provided in Eq. (\ref{wk15visc}) for $N=5$, with the enthalpy factor $\gamma_{0,k}$ being the same for all species:
\begin{equation}
\eta=\frac{T\gamma_0}{10\sigma}\left(x_1C_{0,1}+x_2C_{0,2}+x_3C_{0,3}+x_4C_{0,4}+x_5C_{0,5} \right)
\end{equation}
where $C_{0,1},~C_{0,2}$ are readily defined in Section \ref{iterative}. Iterating $\gamma_{0}x_k=\sum_{l=1}^N C_{0,l}C^{00}_{lk}$ for $N=3,4,5$ is seen as follows: 
\begin{equation}
\begin{split}
\gamma_{0}x_k&=C_{0,1}C^{00}_{1k}+C_{0,2}C^{00}_{2k}+C_{0,3}C^{00}_{3k}~~~~~~~~~~~~~~~~~~~~~~~~~~~~~~~~~(N=3)\\[2ex]
&=C_{0,1}C^{00}_{1k}+C_{0,2}C^{00}_{2k}+C_{0,3}C^{00}_{3k}+C_{0,4}C^{00}_{4k}~~~~~~~~~~~~~~~~~~~(N=4)\\[2ex]
&=C_{0,1}C^{00}_{1k}+C_{0,2}C^{00}_{2k}+C_{0,3}C^{00}_{3k}+C_{0,4}C^{00}_{4k}+C_{0,5}C^{00}_{5k}~~~~~(N=5).
\end{split}
\end{equation}

\subsubsection{$N=3$}

 For $N=3$, solve for $C_{0,3}$ in terms $C^{00}_{3k},~C^{00}_{2k},~C^{00}_{1k},~\gamma_{0},~x_k$, and the previously obtained $C_{0,1},~C_{0,2}$:
\begin{equation}
\begin{split}
&\gamma_{0}x_k=C_{0,1}C^{00}_{1k}+C_{0,2}C^{00}_{2k}+C_{0,3}C^{00}_{3k}\\
&\rightarrow~~~C_{0,3}=\frac{\gamma_0}{C_{3k}^{00}}\left[x_{k}-\frac{x_1C^{00}_{1k}}{C^{00}_{11}}-\frac{C^{00}_{2k}}{C^{00}_{22}}\left(x_2-\frac{2x_1C^{00}_{12}}{C^{00}_{11}} \right) \right].
\end{split}
\end{equation}
Then iterate for $k=1,2,3$:
\begin{equation}
\begin{split}
&(k=1):~~~C^{(1)}_{0,3}=-\frac{\gamma_0C^{00}_{21}}{C_{31}^{00}C^{00}_{22}}\left(x_2-\frac{2x_1C^{00}_{12}}{C^{00}_{11}} \right) \\[2ex]
&(k=2):~~~C^{(2)}_{0,3}=\frac{\gamma_0}{2C_{32}^{00}}\left[x_{2}-\frac{2x_1C^{00}_{12}}{C^{00}_{11}}-\left(x_2-\frac{2x_1C^{00}_{12}}{C^{00}_{11}} \right) \right]=0\\[2ex]
&(k=3):~~~C^{(3)}_{0,3}=\frac{\gamma_0}{C_{33}^{00}}\left[x_{3}-\frac{2x_1C^{00}_{13}}{C^{00}_{11}}-\frac{2C^{00}_{23}}{C^{00}_{22}}\left(x_2-\frac{2x_1C^{00}_{12}}{C^{00}_{11}} \right) \right].
\end{split}
\end{equation}
Add together the nonzero iterations to obtain $C_{0,3}$:
\begin{equation}
\begin{split}
C_{0,3}\Rightarrow \gamma_0\left[\frac{x_3}{C^{00}_{33}}-\frac{2x_1C^{00}_{13}}{C^{00}_{33}C^{00}_{11}}-\left(\frac{x_2}{C^{00}_{22}}-\frac{2x_1C^{00}_{12}}{C^{00}_{22}C^{00}_{11}} \right) \left(\frac{2C^{00}_{23}}{C^{00}_{33}}+\frac{C^{00}_{12}}{C^{00}_{13}} \right)  \right].
\end{split}
\end{equation}

\subsubsection{$N=4$}

 For $N=4$, solve for $C_{0,4}$ in terms $C^{00}_{4k},~C^{00}_{3k},~C^{00}_{2k},~C^{00}_{1k},~\gamma_{0},~x_k$, and the previously obtained $C_{0,1},~C_{0,2},~C_{0,3}$ (leaving $C_{0,3}$ generally defined for the sake of brevity):
\begin{equation}
\begin{split}
\rightarrow~~~C_{0,4}=\frac{\gamma_0}{C_{4k}^{00}}\left[x_{k}-\frac{x_1C^{00}_{1k}}{C^{00}_{11}}-\frac{C^{00}_{2k}}{C^{00}_{22}}\left(x_2-\frac{2x_1C^{00}_{12}}{C^{00}_{11}} \right)-\frac{C_{0,3}C^{00}_{3k}}{\gamma_0} \right].
\end{split}
\end{equation}
Then iterate for $k=1,2,3,4$:
\begin{equation}
\begin{split}
&(k=1):~~~C^{(1)}_{0,4}=\frac{\gamma_0}{2C_{41}^{00}}\left[-\frac{2C^{00}_{21}}{C^{00}_{22}}\left(x_2-\frac{2x_1C^{00}_{12}}{C^{00}_{11}} \right)-\frac{2C_{0,3}C^{00}_{31}}{\gamma_0} \right]\\[2ex]
&(k=2):~~~C^{(2)}_{0,4}=-\frac{C_{0,3}C^{00}_{32}}{C_{42}^{00}}\\[2ex]
&(k=3):~~~C^{(3)}_{0,4}=\frac{\gamma_0}{2C_{43}^{00}}\left[ \frac{C^{00}_{12}C^{00}_{33}}{C^{00}_{22}C^{00}_{13}}\left(x_2-\frac{2x_1C^{00}_{12}}{C^{00}_{11}} \right) \right]\\[2ex]
&(k=4):~~~C^{(4)}_{0,4}=\frac{\gamma_0}{C_{44}^{00}}\left\{x_{4}-2\left[\frac{x_1C^{00}_{14}}{C^{00}_{11}}+\frac{C^{00}_{24}}{C^{00}_{22}}\left(x_2-\frac{2x_1C^{00}_{12}}{C^{00}_{11}} \right)+\frac{C_{0,3}C^{00}_{34}}{\gamma_0}\right] \right\}.
\end{split}
\end{equation}
Add all the iterations together to obtain $C_{0,4}$:
\begin{equation}
\begin{split}
C_{0,4}=C_{0,4}^{(1)}+C_{0,4}^{(2)}+C_{0,4}^{(3)}+C_{0,4}^{(4)}.
\end{split}
\end{equation}

\subsubsection{$N=5$}

 For $N=5$, solve for $C_{0,5}$ in terms $C^{00}_{5k},~C^{00}_{4k},~C^{00}_{3k},~C^{00}_{2k},~C^{00}_{1k},~\gamma_{0},~x_k$, and the previously obtained $C_{0,1},~C_{0,2},~C_{0,3},~C_{0,4}$ (leaving $C_{0,3},~C_{0,4}$ generally defined for the sake of brevity):
\begin{equation}
\begin{split}
\rightarrow~~~C_{0,5}=\frac{\gamma_0}{C_{5k}^{00}}\left[x_{k}-\frac{x_1C^{00}_{1k}}{C^{00}_{11}}-\frac{C^{00}_{2k}}{C^{00}_{22}}\left(x_2-\frac{2x_1C^{00}_{12}}{C^{00}_{11}} \right)-\frac{C_{0,3}C^{00}_{3k}}{\gamma_0}-\frac{C_{0,4}C^{00}_{4k}}{\gamma_0} \right].
\end{split}
\end{equation}
Then iterate for $k=1,2,3,4,5$:
\begin{equation}
\begin{split}
&(k=1):~~~C^{(1)}_{0,5}=\frac{\gamma_0}{2C_{51}^{00}}\left[-\frac{2C^{00}_{21}}{C^{00}_{22}}\left(x_2-\frac{2x_1C^{00}_{12}}{C^{00}_{11}} \right)-\frac{2C_{0,3}C^{00}_{31}}{\gamma_0}-\frac{2C_{0,4}C^{00}_{41}}{\gamma_0} \right]\\[2ex]
&(k=2):~~~C^{(2)}_{0,5}=\frac{-1}{C^{00}_{52}}\left(C_{0,3}C^{00}_{32}+C_{0,4}C^{00}_{42}\right)\\[2ex]
&(k=3):~~~C^{(3)}_{0,5}=\frac{-C_{0,4}C^{00}_{43}}{C_{53}^{00}}\\[2ex]
&(k=4):~~~C^{(4)}_{0,5}=\frac{\gamma_0}{2C_{54}^{00}}\Big(x_{4}-\frac{C_{0,4}C^{00}_{44}}{\gamma_0}-2\Big[\frac{x_1C^{00}_{14}}{C^{00}_{11}}+\left(x_2-\frac{2C^{00}_{12}}{C^{00}_{11}} \right)C^{00}_{24}\\
&~~~~~~~~~~~~~~~~~~~~~~~~~~~~~~~~~~~~~~~~~~~~~~~~~~~~~~~~~+\frac{C_{0,3}C^{00}_{34}}{\gamma_0}\Big] \Big)\\
&(k=5):~~~C^{(5)}_{0,5}=\frac{\gamma_0}{C_{55}^{00}}\Big\{x_{5}-2\Big[\frac{x_1C^{00}_{15}}{C^{00}_{11}}+\frac{C^{00}_{25}}{C^{00}_{22}}\left(x_2-\frac{2x_1C^{00}_{12}}{C^{00}_{11}} \right)\\
&~~~~~~~~~~~~~~~~~~~~~~~~~~~~~~~~~~~~~~~~~~+\frac{C_{0,3}C^{00}_{35}}{\gamma_0}+\frac{C_{0,4}C^{00}_{45}}{\gamma_0}\Big] \Big\}.
\end{split}
\end{equation}
All the iterations are combined to obtain $C_{0,5}$:
\begin{equation}
\begin{split}
C_{0,5}=C_{0,5}^{(1)}+C_{0,5}^{(2)}+C_{0,5}^{(3)}+C_{0,5}^{(4)}+C_{0,5}^{(5)}.
\end{split}
\end{equation}

\subsection{Application: An $N\leq5$ QGP} \label{5qgp}

Appendix A in Ref. \cite{Arnold:2003zc} displays a table of $|\mathcal{M}|^2/g^4$ scattering amplitudes for relevant parton interactions in a QGP. Certain elastic interactions have the same amplitude, which means that these interactions have the same differential cross section. In an $N=5$ QGP with the parton species numbered as follows: $g=1,~q_1=2,~q_2=3,~\bar{q}_1=4,~\bar{q}_2=5$, we have the following elastic interactions that have distinctive and/or identical $|\mathcal{M}|^2/g^4$:

\begin{table}[h]
\caption{Elastic $l+k\rightarrow l+k$ interactions in pQCD that have distinctive and/or identical $|\mathcal{M}|^2/g^4$}\label{tab5}
\begin{tabular}{l|lllll}
\toprule
Parton Interactions & $gg$ & $gq_1$ & $q_1q_1$ & $q_1q_2$  & $q_1\bar{q}_1$  \\
& & $g{q}_2$ & $q_2q_2$ & $q_1\bar{q}_2$  & ${q}_2\bar{q}_2$  \\
& & $g\bar{q}_1$ & $\bar{q}_1\bar{q}_1$ & $\bar{q}_1q_2$  &   \\
& & $g\bar{q}_2$ & $\bar{q}_2\bar{q}_2$ & $\bar{q}_1\bar{q}_2$  &   \\
\midrule
$l,k$ Index Labels for $C^{00}_{lk}$ &11 & 12,~21 & 22 & 23,~32  & 24,~42\\
 
&  & 13,~31 & 33 & 34,~43 & 35,~53 \\
 
& & 14,~41 & 44 &45,~54 & \\
 
& & 15,~51 & 55 & 52,~25& \\
\midrule
Simplified Index Labels&11 & 12 & 22 & 23  & 24\\
\botrule
\end{tabular}
\end{table}

The last row of Table \ref{tab5} is the simplification of the interaction numbering for the collision kernels $C^{00}_{lk}$, which denotes a specific elastic QCD interaction: 11 is the gluon--gluon interaction; 12 is the gluon--fermion interaction; 22 is the like-fermion interaction; 23 is the different-fermion interaction; and 24 is the same-flavor quark/anti-quark interaction.

As the species types 2, 3, 4, and 5 are related to (anti-)quarks, they have the same molar fraction via an identical quantum number density, i.e., $x_2=x_3=x_4=x_5$. This vastly simplifies, in most cases, the $C_{0,l}$ factors to the following:
\begin{equation}
\begin{split}
&C_{0,1}=\frac{\gamma_{0}x_1}{C^{00}_{11}}\\\\
&C_{0,2}=\frac{\gamma_0}{C^{00}_{22}}\left(x_{2}-\frac{2x_1C^{00}_{12}}{C^{00}_{11}} \right)\\\\
&C_{0,3}= C_{0,2}\left[\frac{-2C^{00}_{23}}{C^{00}_{22}}\right]\\\\
&C_{0,4}=C_{0,2} \left[ \frac{2C^{00}_{23}}{C^{00}_{22}}\left(1 +\frac{C^{00}_{23}}{C_{24}^{00}}+\frac{2C^{00}_{23}}{C^{00}_{22}} \right)+\frac{C^{00}_{22}}{2C^{00}_{23}}-\frac{2C^{00}_{24}}{C^{00}_{22}}\right]\\\\
&C_{0,5}=C_{0,2}\Big[-1-\frac{C^{00}_{22}}{2C_{24}^{00}}-\frac{C^{00}_{22}}{2C^{00}_{23}}\left(1+\frac{C^{00}_{22}}{2C^{00}_{23}}\right)+\frac{2C^{00}_{23}}{C_{22}^{00}}\left(1+\frac{4C^{00}_{24}}{C_{22}^{00}}\right)\\
&~~~~~~~~~~~~~~~~~+\left(1 -\frac{C^{00}_{23}}{C_{24}^{00}}-\frac{2C^{00}_{23}}{C^{00}_{22}} \right)\left(\frac{2C^{00}_{23}}{C^{00}_{22}}+ \frac{2C^{00}_{24}}{C^{00}_{22}}\right)\\
&~~~~~~~~~~~~~~~~~-\left(1 +\frac{C^{00}_{23}}{C_{24}^{00}}+\frac{2C^{00}_{23}}{C^{00}_{22}} \right)\left(1+\frac{2C^{00}_{23}}{C^{00}_{22}}\frac{C^{00}_{23}}{C_{24}^{00}}+ \frac{4C^{00}_{23}C^{00}_{23}}{C^{00}_{22}C_{22}^{00}} \right) \Big].
\end{split}
\end{equation}
The (anti-)quark factors, $C_{0,3}$ through $C_{0,5}$, depend on the quark--gluon factor $C_{0,2}$.

In a hypothetical case where gluons are neglected, such that only (anti-)quark matter in a quasifree phase can be analyzed, $C_{0,1}=0$ naturally and $C_{0,2}=x_2\gamma_0/C^{00}_{22}$. This leads to the (anti-)quark factors being dependent only on the like-quark factor.

 In another hypothetical case where anti-quarks should be ignored in an effective $N=3$ QGP of different-flavor quarks and gluons, the factors that involve anti-quarks naturally vanish: $C_{0,4}=C_{0,5}=0$.

\section{Conclusion}

In this study, I offer a variation of the Chapman--Enskog viscosity formula for an $N$-component mixture. There are two elements to this variation: how the collision kernel linearized via the Ritz variational method can be approximated for an ultrarelativistic gas, i.e., massless gas under anisotropic scatterings, and how the shear viscosity of gaseous mixtures of species numbers beyond $N=2$ can be alternatively defined through iteration.

Significant findings were made and reported in Sections \ref{ritzy} and \ref{iterative}. In Section \ref{ritzy}, the approximated linearized collision kernel is expressed in terms of the relativistic omega integrals and vanishing terms that depend on species type and total number of species. The simpler scenario of isotropic binary mixtures is utilized for effective comparison with the well-established Moroz framework, where the kernels were compared side-by-side via a 1:1 ratio in Section \ref{geniso}. Comparison between the approximated and Moroz kernels extended into Section \ref{iterative}, as an iterative formula for CE viscosity is compared with the conventional formula for a binary mixture -- using the $N=2$ QGP as a specific example. Agreement between the conventional and iterative formulas agreed well for noninteracting systems, and moderately for interacting systems.

Future endeavors branching from this study would include a proper analysis of an $N\leq5$ QGP, where the viscosity factors are specifically defined for gluons and (anti-)quarks in Section \ref{5qgp}. Rather than forcing parton interactions to be isotropic, which would not make physical sense in the pQCD regime, we can properly utilize the forward-peaking differential cross sections of parton interactions. Therefore, one can use the approximated kernels and iterative formula offered in this report.

\begin{appendices}

\section{Context: The Green--Kubo Relation}

Numerical methods, such as the Green--Kubo relation \cite{Green:1954ubq, Kubo:1957mj, Muronga:2003tb, Demir:2008tr, Fuini:2010xz}, use correlation functions of the volume-averaged, off-diagonal terms of the energy-momentum tensor, $\bar{\pi}^{xy}$, between the initial and final times $t_0$ and $t$. This would then be averaged over the final time $t$, to calculate $\eta$ in equilibrium:
\begin{equation}
\eta= \frac{V}{T}\int_0^\infty \langle\bar{\pi}^{xy}(t+t_0)\bar{\pi}^{xy}(t_0)\rangle dt.
\end{equation}
In the above, the volume of the system $V$ is factored with the time average to remove the volume density from the extracted data, similar to how the time is integrated all over to remove any time dependency. The numerical calculations are strictly dependent on the off-diagonal stress tensor, which can be compared with analytical methods such as the Chapman--Enskog method. Given that these correlation functions dampen exponentially over time \cite{Muronga:2003tb, Demir:2008tr, Fuini:2010xz}, the Green--Kubo viscosity takes on a semi-analytical, relaxation time profile \cite{Wang:2021owa}: 
\begin{equation}
\eta=\frac{4}{15}\epsilon\tau.
\end{equation}

In the above, $\epsilon$ is the energy density of the partons in equilibrium, and $\tau$ is the characteristic relaxation time of the gas. In the Green--Kubo context, the relaxation time is extracted through numerical calculations. For the case of a QGP, numerical calculations are computed via parton cascade models such as the ZPC model \cite{Zhang:1997ej}. Previous studies, as mentioned in the introduction, have shown that Green--Kubo calculations under the ZPC model mutually agree with the Chapman--Enskog formula using a scattering cross section based on the pQCD gluon--gluon amplitude \cite{Wang:2021owa, MacKay:2022uxo}.

\section{pQCD-based Differential Cross Sections}

For an elastic collision between two (effectively) massless particles in the CM frame, the differential cross section $d\sigma/d\Omega$ is directly proportional to the square of the scattering amplitude of the given interaction, $|\mathcal{M}|^2$ \cite{griffiths}. Based on the $\hat{s}$- and $\hat{t}$-Mandelstam variables, the differential cross section is inversely proportional to $\hat{s}^2$, as it is differentiated with respect to the momentum transfer variable $\hat{t}$:
\begin{equation}
\frac{d\sigma}{d\hat{t}}=\frac{1}{16\pi\hat{s}^2}\langle|\mathcal{M}|^2\rangle.
\end{equation} 
If the interacting particles have mass $m_1=m_2=m$, the denominator factor would instead be $16\pi \hat{s}(\hat{s}-4m^2)$.

In real-world quantum field theories, the square of the scattering amplitude is the sum of all spin states of the external fermion and boson lines of a given Feynman diagram, as per Casimir's Trick. To remove any spin overcounting from the $|\mathcal{M}|^2$ calculation, the spin states of the initial particles are averaged. For QCD diagrams, the color states of the fermion/gluon lines are also summed and averaged. In Appendix A of Ref. \cite{Arnold:2003zc}, the provided square of the amplitudes are summed over all the spin-color states; however, they are not averaged. From these amplitudes, one can obtain differential cross sections where the total cross section is energy independent, applicable for the ``large $\hat{s}$'' case where one drives $\sqrt{\hat{s}}\rightarrow\infty$.

\subsection{Gluon--Gluon}

The gluon--gluon scattering amplitude is provided as follows (also to be seen in Ref. \cite{schwartz}):
\begin{equation}
|\mathcal{M}|^2=1152g^4\left(3-\frac{\hat{t}\hat{u}}{\hat{s}^2}-\frac{\hat{s}\hat{u}}{\hat{t}^2}-\frac{\hat{s}\hat{t}}{\hat{u}^2}\right).
\end{equation}
Here, $\hat{u}$ is the Mandelstam variable for momentum transfer where external outgoing lines are crossed. For the case of massless particles, which indeed applies for gluons and for ultrarelativistic particles, $\hat{u}=-\hat{s}-\hat{t}$. As gluons have 2 spin states and 8 gluon numbers, the spin-color averaging factor is $(2\times8)^2=256$, where the average square of the magnitude is
\begin{equation}
\langle|\mathcal{M}|^2\rangle=\frac{9g^4}{2}\left(3-\frac{\hat{t}\hat{u}}{\hat{s}^2}-\frac{\hat{s}\hat{u}}{\hat{t}^2}-\frac{\hat{s}\hat{t}}{\hat{u}^2}\right).
\end{equation}

For the perturbative case where the momentum transfer is very small, i.e., CM energy is very large, $\langle|\mathcal{M}|^2\rangle$ can be revised where only the divergent parts remain:
\begin{equation}
\lim_{\hat{s}\rightarrow\infty}\langle|\mathcal{M}|^2\rangle=\frac{9g^4}{2}\frac{\hat{s}^2}{\hat{t}^2}.
\end{equation}
To remove any divergence from the total cross section, the Debye screening mass $m_D$ is introduced to the $\hat{t}$-channel (c.f. Appendix A.1 ``\textit{Gauge boson self energy}'' in Ref. \cite{Arnold:2003zc}):
\begin{equation}
\langle|\mathcal{M}|^2\rangle=\frac{9g^4}{2}\frac{\hat{s}^2}{(\hat{t}-m_D^2)^2}.
\end{equation}
This equation is inserted into the definition for $d\sigma/d\hat{t}$, where the total cross section is evaluated over $\hat{t}\in[-s,0]$ with $\hat{s}\rightarrow\infty$:
\begin{equation}
\frac{d\sigma}{d\hat{t}}=\frac{9g^4}{32\pi}\frac{1}{(\hat{t}-m_D^2)^2},~~~\Rightarrow~~~\sigma=\int_{-\infty}^0\frac{d\sigma}{d\hat{t}}d\hat{t}=\frac{9g^4}{32\pi m_D^2}.
\end{equation}

If $\sigma$ is obtained while allowing $\hat{s}$ to be present, the integral is evaluated as energy dependent. To make the total cross section energy independent, i.e., independent of $\hat{s}$, the energy dependent factor is scaled away from the total cross section by factoring in its reciprocal. The reciprocal of this factor is included in the gluon--gluon differential cross section, such that we have a redefined gluon--gluon $d\sigma/d\hat{t}$ that defines an energy independent total cross section:
\begin{equation}
\begin{split}
\sigma=\int_{-\hat{s}}^0\frac{d\sigma}{d\hat{t}}d\hat{t}=\frac{9g^4}{32\pi}&\left(1+\frac{m_D^2}{\hat{s}} \right)^{-1},\\
&\Rightarrow~~~\frac{d\sigma'}{d\hat{t}}=\frac{9g^4}{32\pi}\left(1+\frac{m_D^2}{\hat{s}} \right)\frac{1}{(\hat{t}-m_D^2)^2}.
\end{split}
\end{equation}
The redefinition of $d\sigma'/d\hat{t}$ is the differential cross section used in parton cascade and multi-phase transport models, such as the ZPC and AMPT models, that analyze a gluon-dominant QGP with $g^2=4\pi\alpha_s$ \cite{Xu:2008av, Ferini:2008he, Molnar:2001ux, Zhang:1997ej, Wang:2021owa, MacKay:2022uxo, Plumari:2012ep, Lin:2004en}.

\subsection{Quark--Gluon}

The average square of the magnitude for the quark--gluon interaction -- with the spin-color averaging being $(2\times3)\cdot(2\times8)=96$, with the quark mass replaced with a Fermi screening mass $m_F$ and the boson exchange screened with the Debye mass -- is
\begin{equation}
\langle|\mathcal{M}|^2\rangle=\frac{g^4}{3}\left[-\frac{4}{3}\left(\frac{\hat{t}\hat{u}}{(\hat{t}-m_F^2)^2}+\frac{\hat{u}\hat{s}}{(\hat{u}-m_F^2)^2} \right)+3\frac{\hat{s}^2+\hat{u}^2}{(\hat{t}-m_D^2)^2} \right].
\end{equation}
For the perturbative, ultrarelativistic cases of $\hat{u}=-\hat{s}-\hat{t}$ and $\hat{s}\rightarrow\infty$, the averaged square of the magnitude simplifies to
\begin{equation}
\lim_{\hat{s}\rightarrow\infty}\langle|\mathcal{M}|^2\rangle=\frac{6g^4}{3}\frac{\hat{s}^2}{(\hat{t}-m_D^2)^2} .
\end{equation}
This equation is inserted into the definition for $d\sigma/d\hat{t}$ to solve for the total cross section with $\hat{s}\rightarrow\infty$:
\begin{equation}
\frac{d\sigma}{d\hat{t}}=\frac{g^4}{8\pi}\frac{1}{(\hat{t}-m_D^2)^2},~~~\Rightarrow~~~\sigma=\int_{-\infty}^0\frac{d\sigma}{d\hat{t}}d\hat{t}=\frac{g^4}{8\pi m_D^2}.
\end{equation}

\subsection{Quark--Quark}

For quark--quark interactions, it is important to determine whether the quarks have the same flavor or different flavors. For the case of same flavor quarks, with a spin-color average of $(2\times3)^2=36$, the averaged square of the magnitude, using fermion and boson screening masses, is
\begin{equation}
\begin{split}
\langle|\mathcal{M}|^2\rangle=\frac{4g^4}{9}\Big[&\left(\frac{\hat{s}^2+\hat{u}^2}{(\hat{t}-m_D^2)^2}+\frac{\hat{s}^2+\hat{t}^2}{(\hat{u}-m_D^2)^2} \right)\\
&-\frac{2}{3}\left(\frac{\hat{s}\hat{t}}{(\hat{t}-m_F^2)^2}+\frac{\hat{s}\hat{u}}{(\hat{u}-m_F^2)^2} \right)\Big].
\end{split}
\end{equation}
By utilizing $\hat{u}=-\hat{s}-\hat{t}$ and $\hat{s}\rightarrow\infty$, the averaged square of the magnitude simplifies to
\begin{equation}
\lim_{\hat{s}\rightarrow\infty}\langle|\mathcal{M}|^2\rangle=\frac{8g^4}{9}\frac{\hat{s}^2}{(\hat{t}-m_D^2)^2}.
\end{equation}
The total cross section with $\hat{s}\rightarrow\infty$ can therefore be solved:
\begin{equation}
\frac{d\sigma}{d\hat{t}}=\frac{g^4}{18\pi} \frac{1}{(\hat{t}-m_D^2)^2},~~~\Rightarrow~~~\sigma=\int_{-\infty}^0\frac{d\sigma}{d\hat{t}}d\hat{t}=\frac{g^4}{18\pi m_D^2}.
\end{equation}

For the quark--gluon and same-flavor quark differential cross sections, the energy factor $(1+m_D^2/\hat{s})$ can be factored in to redefine the respective $d\sigma/d\hat{t}$ formulas, such that each obtain an energy-independent total cross section over any $\hat{s}$.



\end{appendices}


\end{document}